\documentclass{article}

\usepackage{arxiv}

\usepackage[utf8]{inputenc} 
\usepackage[T1]{fontenc}    
\usepackage{hyperref}       
\usepackage{url}            
\usepackage{booktabs}       
\usepackage{amsfonts}       
\usepackage{amsmath}
\usepackage{amssymb}
\usepackage{nicefrac}       
\usepackage{microtype}      
\usepackage{lipsum}		
\usepackage{graphicx}
\usepackage{color}
\usepackage[numbers]{natbib}
\usepackage{doi}

\pdfoutput=1

\begin{document}


\title{The Area Localized Coupled Model for Analytical Mean Flow Prediction in Arbitrary Wind Farm Geometries} 



\author{Genevieve M. Starke$^1$, Charles Meneveau$^1$, Jennifer R. King$^2$, and Dennice F. Gayme$^1$}

\thanks{$^{1}$ Genevieve M. Starke, Charles Meneveau and Dennice F. Gayme are with the Department of Mechanical Engineering, Johns Hopkins University,
Baltimore, Maryland 21218, USA; email:
        {\url{ gstarke1@jhu.edu, meneveau@jhu.edu, dennice@jhu.edu}}}%
\thanks{$^{2}$ Jennifer King is with the National Renewable Energy Laboratory; email:{\url{ Jennifer.King@nrel.gov}.}}%


\date{\today}
\maketitle

\begin{abstract}

This work introduces the Area Localized Coupled (ALC) model, which extends earlier approaches to coupling classical wake superposition and atmospheric boundary layer models in order to enable applicability to arbitrary wind-farm layouts. Coupling wake and top-down boundary layer models is particularly challenging since the latter requires averaging over planform areas associated with certain turbine-specific regions of the flow. The ALC model uses Voronoi tesselation to define a local area around each turbine. A top-down description of a developing internal boundary layers is then applied over Voronoi cells upstream of each turbine to estimate the local mean velocity profile. Coupling between the velocity at hub-height based on this localized top-down model and a wake model is achieved by enforcing a minimum least-square-error in mean velocity in each cell. The ALC model is implemented using a wake model with a profile that transitions from a top-hat to Gaussian function and accounts for wake interactions through  linear superposition. Detailed comparisons to large-eddy simulation (LES) data demonstrate the efficacy of the model in accurate predictions of both power and hub height velocity for  complex wind farm geometries. Further validation with LES for a hybrid array-random farm that has half of the turbines arranged in an array and the other half randomly distributed indicates the model's versatility with respect to capturing results from different wind farm configurations.  In both cases, the ALC model is shown to produce improved power predictions for both the farm and individual turbines over prevailing approaches for a range of wind inflow directions.


%
\end{abstract}


\maketitle 

\section{Introduction}
Analytical modeling of wind farms has been approached in two main ways: a wake model approach or a top-down boundary layer approach.  The wake model method focuses on wakes of individual turbines and their superposition to predict the mean velocity field of the entire farm.  Extensive work has been done in this area, both on the modeling of the velocity deficit, using different analytical functions\cite{Jensen1983a, Bastankhah2014a, TIAN201590, Annoni2018, Ge2019a} experimental models\cite{Barthelmie2003, Iungo2014} and data-driven methods\cite{Ti2020a}, and on the superposition of the deficits \cite{Niayifar2015a,Niayifar2016a,Shapiro2019,zong_porte_agel2020}  The top-down model method focuses on larger scales and considers the wind farm's interactions with the atmospheric boundary layer (ABL) \cite{Frandsen2006a, Calaf2010a, Abkar}.  While both of these approaches are capable of representing certain aspects of the flow in wind farms, coupling the two approaches promises to leverage the strengths of each to improve the accuracy of the final mean flow prediction in the wind farm.
\begin{figure*}[ht!]
  \begin{center}
    \includegraphics[scale=1]{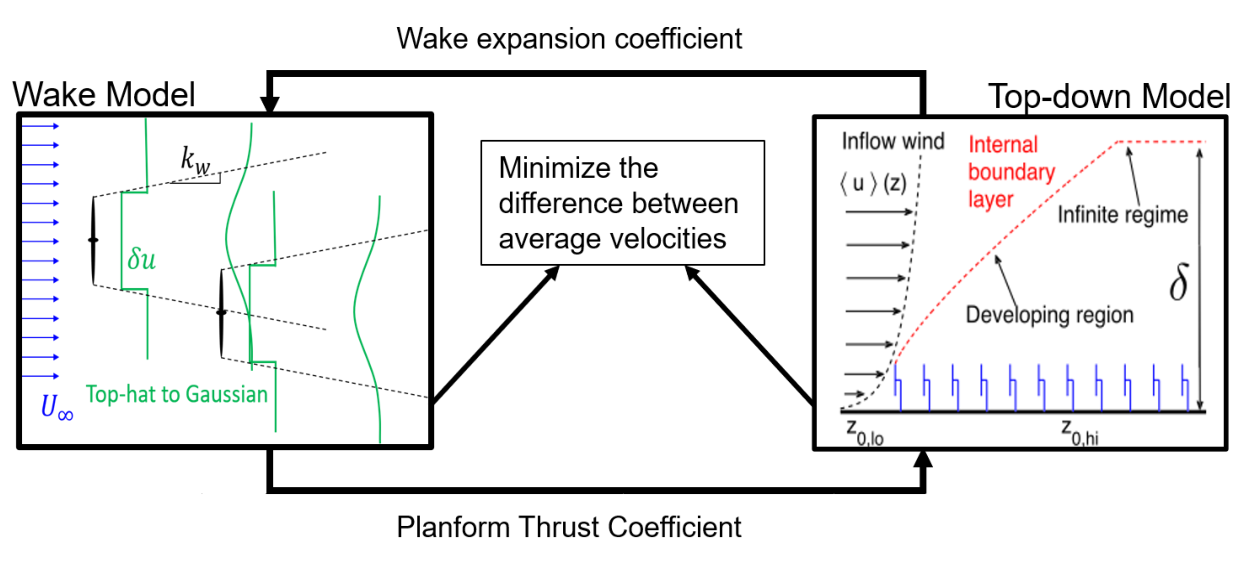}
  \end{center}
  \caption{\label{fig:overview first} Illustration of the ALC model, where the wake model and top-down model are connected with the wake expansion coefficient and the planform thrust coefficient.}
\end{figure*}

The first coupled model to be proposed was by Frandsen\cite{Frandsen2006a}, who also proposed one of the first top-down models.  The Frandsen model includes three regimes, the first of which is located at the beginning of the farm, followed by the second and third regimes further back in the farm.  In the first regime, a wake model is applied that uses a particular set of assumptions regarding wake growth and wake velocity defect superposition.  The third regime, far downstream, is where the wind farm is considered to have reached a fully developed state (similar to an infinite wind farm), and is in balance with the ABL, i.e. a top-down model.  This idea sparked other top-down models, some building upon the Frandsen model, such as the model of Calaf et al.\cite{Calaf2010a} including a more detailed description of the mean velocity profile in the turbine layer. Top down models have also been extended to include various atmospheric conditions \cite{Pena2013a,sescu2015large}.  

While the Frandsen coupled model proposed a novel way of coupling the wake model and a top-down model, it was devised for a single column of turbines in a farm composed of many columns of turbines.  This formulation limited its application mostly to regular arrays and made applications to arbitrary wind farm geometries a challenge.  

The Coupled Wake Boundary Layer (CWBL) Model, proposed by Stevens et al\cite{Stevens2015a,Stevens2016c}, widens the applicability of coupled models by using a top-down model that had been modified for application to finite-length wind farms \cite{Meneveau2012a}.  This top-down model only required a `fully-developed' region in the wind farm, which is an area of the wind farm where the boundary layer created by the turbines is no longer growing.  This enabled its application to wind farms long enough to contain this region.  The CWBL model uses the Jensen model\cite{Jensen1983a,Katic1986a} for its wake model component, which is one of the most used wake models in applications.  The Jensen model uses a top-hat velocity profile and a square superposition of wake velocity deficits.  This model was applied to data from large eddy simulations (LES) and existing wind farms with good results.  However, this model coupled the wake model and the top-down model using constants derived from the streamwise and spanwise spacing of the wind turbines, confining its application to uniform wind farms. The more general version of the model\cite{Stevens2016c} required the identification of a fully developed region, assuming that it exists within the wind farm. Applications to non-regular turbine arrangements remained a challenge. 

As more wind farms are being built, especially on-shore ones where topographical features often dictate irregular wind turbine arrangements, there is a need for a more generalizable model that can be applied to any wind farm regardless of turbine arrangement.  The model by Shapiro et al\cite{Shapiro2019} localized the top-down model to each turbine instead of involving averaging over an entire region of the wind farm. This allowed applications to general wind farm layouts and include a description of the growth of internal boundary layers for each turbine instead of being limited to the fully developed region only.  The Shapiro et al\cite{Shapiro2019}  model uses a more realistic wake profile than the usual Jensen model, representing the deficit behind the turbine as a super Gaussian profile that first approximates a top-hat profile near the turbine, and then transitions smoothly towards a Gaussian profile\cite{Bastankhah2014a,Shapiro2019,Blondel20a} further downstream. The approach also used a linear superposition of wake velocity deficits.
This method was compared with LES results\cite{Shapiro2019} for a uniform wind farm and showed an improvement over the CWBL Model in several wind directions.  This model can be applied to a nonuniform wind farm, but still uses a uniform inflow.  This becomes a limiting assumption as wind farms cover larger areas where the flow conditions can change across the wind farm.

In this paper, we build upon these coupled models and present the Area Localized Coupled (ALC) model, which enables the application to wind farms with irregularly spaced turbines. We also include the possibility of a nonuniform inflow velocity profile.  The ALC model combines a wake model that uses the super-Gaussian wake approach and a linear wake superposition\cite{Shapiro2019} with the top-down model of a wind farm in the atmospheric boundary layer\cite{Calaf2010a,Shapiro2019}.  
The wake model uses a super-Gaussian for the wake profile to better match experimental data of wind turbine wake behavior.  The top-down model provides a larger scale picture, using an individual turbine formulation along with a more collective developing boundary layer description of the flow.  The models are connected in an iterative loop (see Fig. \ref{fig:overview first}), where matching the combined descriptions is used to determine the wake expansion coefficients for the wake model as well as the planform thrust coefficient for the top-down model.
\par
An overview of the model is presented in Section \ref{sec2} with brief descriptions of both the wake top-down models that constitute the building blocks for the ALC model.  In Section \ref{sec3}, we validate the model using LES data from a circular wind farm.  In Section \ref{sec4}, we compare the model results with LES data from a wind farm that contains two regions, one with a regular array and another with a random distribution of turbines.  Finally, in Section \ref{Sec5}, we present conclusions from our work.

\section{\label{sec2} Area Localized Coupled (ALC) Model}
The following sections summarize the two reduced-order models that are combined in the ALC model, the wake model and the top-down model.

\subsection{Wake Model}
The aim of the wake model is to calculate a velocity field using the aggregated wakes of the turbines in the farm.  The flow field is described by the equation \cite{Shapiro2019}
\begin{equation}
    u(x,y,z) = U_\infty (y) - \sum_n \delta u_n(x) \, W_n(x',r').
    \label{eq:velwake}
\end{equation}
where $x$ is the streamwise coordinate (in the direction of the incoming freestream wind), $x'=x-s_{n,x}$ is the coordinate relative to the position of the turbine, $s_{n,x}$, $r'=[(y-y_n)^2+(z-z_n)^2]^{1/2}$ is the radial distance from the center position of the $n^{\text{th}}$ turbine, $U_{\infty}(y)$ is the incoming freestream mean velocity at hub-height, which can vary across the inflow plane, $y$ is the horizontal coordinate transverse to the incoming wind, $z$ is the vertical coordinate, $\delta u_n(x)$ is the deficit velocity of the $n^{\text{th}}$ turbine at downstream position $x$, and $W_n(x,r)$ is each wake's shape function.  For this model, the velocity deficit $\delta u_n(x)$ is calculated using the steady-state solution to a one-dimensional partial differential equation that describes the behavior of the wake deficit velocity based on the wake expansion and a deficit forcing term.  The latter is prescribed to impose the correct initial velocity deficit just downstream of each turbine.  The steady-state solution was derived in Shapiro et al \cite{Shapiro2019} and has the form
\begin{equation}
    \delta u_n(x) = \frac{\delta u_{0,n}}{2(d_{w,n}(x))^2} \left[1 + \text{erf}\left( \frac{x}{\Delta \sqrt{2}}\right)  \right],
    \label{eq:steadusol}
\end{equation}
where $\delta u_{0,n}$ is the initial velocity deficit, $d_{w,n}(x)$ is the wake expansion function, and $\Delta$ is the characteristic width of the Gaussian function applied to the forcing in the partial differential equation.
The wake expansion function has the form 
\begin{equation}
    d_{w,n}(x) = 1 + k_w ln(1 + e^{x/R}),
\end{equation}
where the exponential was added to prevent the natural log from being undefined and $k_w$ is the wake expansion coefficient, which determines the rate at which the wake expands as it travels downstream.  The expansion of a wake depends on local flow conditions and turbulence, and is defined as 
\begin{equation}\label{eq:k}
    k_{w,n} = \alpha \, \frac{u_{*,n}}{u_{\infty,n}}.
\end{equation}
It depends on the local friction velocity $u_{*,n}$ which is derived from the top-down model and $\alpha$, which is left as a model parameter yet to be determined. We describe how to determine the local friction velocity $u_{*,n}$ valid for a particular turbine $n$ in the next section.  
\par
The initial velocity deficit is obtained from an inviscid model and has the traditional form based on actuator disk momentum theory:
\begin{equation}
   \delta u_{0,n} = u_{\infty,n} (1 - \sqrt{1 - C_{T,n}}) = 1/2 C_{T,n}\prime u_{d,n},
\end{equation}
as described in Shapiro et al \cite{Shapiro2019}.  The nonuniform inflow also affects the velocity deficit calculations by determining the ``upstream'' velocity $u_{\infty,n}$ of each turbine.  To calculate $u_{\infty,n}$, i.e. the disk velocity that would have existed at wind turbine $n$ without the turbine there, we use the following expression:

\begin{equation}
         u_{\infty,n} =  \int_0^{R} \left[U_{\infty}(y) - \sum_{m\neq n} \delta u_m (s_{n,x}) \, W_m(s_{n,x}-s_{m,x},r_n) \right] 2 \pi r_n \ d r_n,
\end{equation}

where $R$ is the radius of the turbine, $s_{n,x}$ is, again, the position of the turbine in the streamwise direction and $r_n =[(y-y_n)^2+(z-z_n)^2]^{1/2}$ is the radial coordinate of point $(y,z)$ for turbine $n$ with rotor center at $(y_n,z_n)$.  For practical purposes, this quantity, and the deficits, are calculated from the front of the farm to the back, in regards to the current wind direction, and the sum is performed over all turbines that have already been calculated.  The first turbines in the farm  use the freestream velocity: $u_{\infty,n}=U_{\infty}(y_n)$ for $n$ where turbines are exposed to the freestream.   
\par
The wake deficits are superposed using the weighting provided by the wake shape function $W_n$.  This function assumes an initial top-hat wake profile that transitions into a Gaussian wake profile downstream.  The wake function is proposed to have the form of a super-Gaussian and defines the wake 
as
\begin{equation}
    W(x,r) = C(x) \, \text{exp}\left(\frac{-D^2}{8\sigma_0^2} \left( \frac{2r}{D \ d_w(x)}\right)^{p(x)} \right)
\end{equation}
where $\sigma_0^2=D/4$ is a parameter that defines the width of the wake relative to the wake diameter and $r^2=y^2+z^2$. The exponent $p(x)$ determines the shape and its dependence on $x$ dictates how fast the transition from a top-hat to a Gaussian occurs. The following functional form was proposed\cite{Shapiro2019} to reproduce experimental wake data:  
\begin{equation}
    p(x) = 2\left(1 + \frac{D}{\text{max}(x,0)} \right).
\end{equation}
The function $C(x)$ can then be found to have the following form assuming  mass conservation (or linearized momentum conservation), as shown in \cite{Shapiro2019},
\begin{equation}
    C(x) = \frac{p(x)}{2\Gamma \left(2/p(x) \right)} \left( \frac{D^2}{8\sigma_0^2} \right)^{2/p(x)}.
\end{equation}
We remark that Eq. \ref{eq:steadusol} is the steady state solution to the following   partial differential equation that describes the one-dimensional time-dependent behavior of the wake:
\begin{eqnarray}
    \frac{\partial \delta u_n}{\partial t} + U_{\infty} \frac{\partial \delta u_n}{\partial x} = - & & \frac{2U_{\infty}}{(d_{w,n}(x))^2} \frac{d(d_{w,n})}{d x} \delta u_n (x,t) + \nonumber \\
    & & \left[U_{\infty} \delta u_{0,n}\right] \left[ \frac{1}{\Delta \sqrt{2 \pi}} \text{exp}\left( \frac{-x^2}{2 \Delta^2} \right) \right].
\end{eqnarray}
In the present work, we do not take into account the time-dependence. However, with $\delta u_n(x,t)$ determined from this PDE, the ALC model can be readily extended to time-varying applications. Returning to the steady state case, the wake model is now complete as long as a value for the parameter $\alpha$ can be specified unambiguously. To accomplish that requires matching with the top-down model described in the next section. 

\begin{figure*}[ht!]
  \begin{center}
    \includegraphics[scale=0.85]{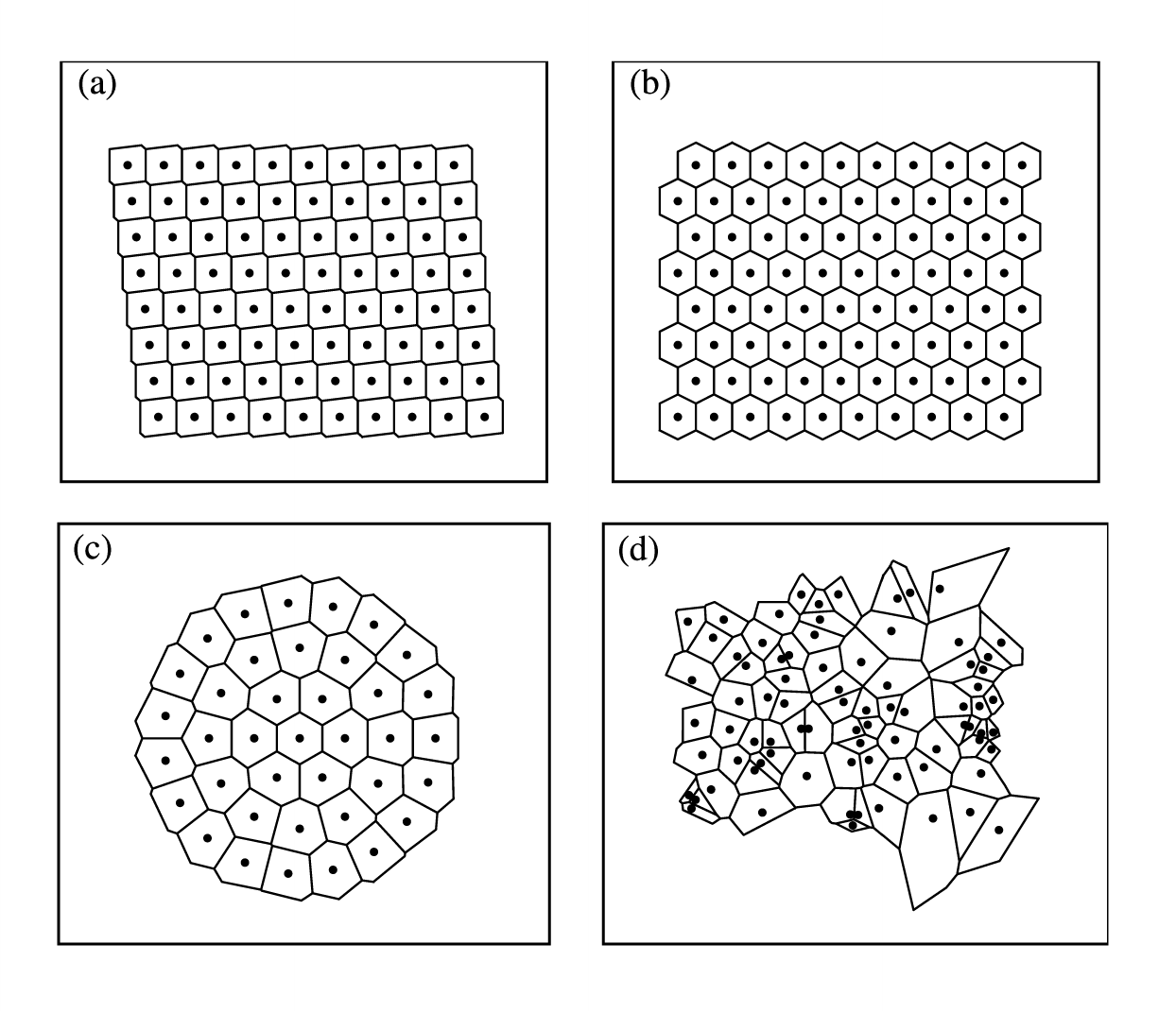}
  \end{center}
  \caption{\label{fig:voronoi} Voronoi cells drawn for different farm configurations including (a) a uniform farm, (b) a staggered farm, (c) a circular farm and (d) a random farm}
\end{figure*}

\begin{figure*}[ht!]
  \begin{center}
    \includegraphics[scale=0.8]{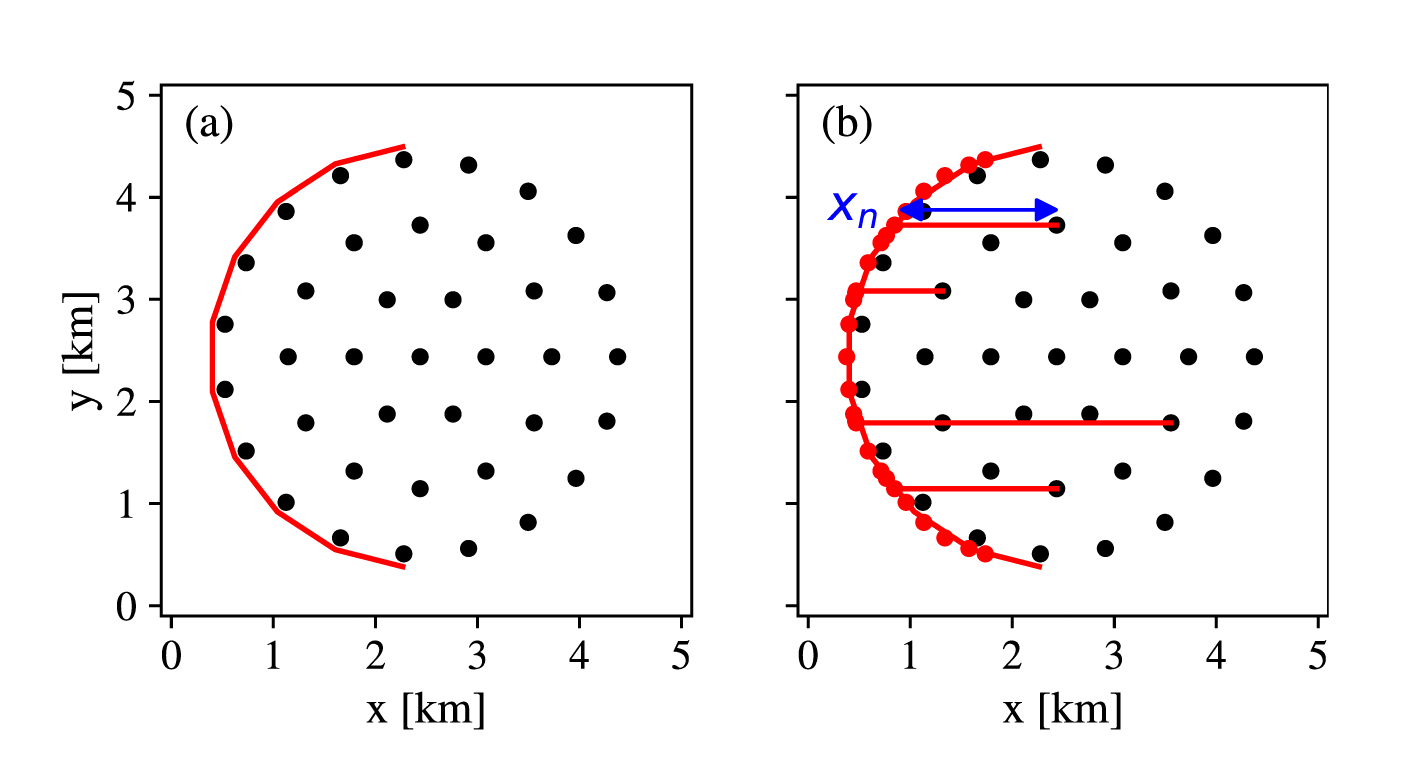}
  \end{center}
  \caption{\label{fig:BL} (a) The new definition of the start of the boundary layer for the top-down model starting in front of the first turbines and (b) how the $x_n$ parameter is found for each turbine.}
\end{figure*}

\subsection{Top-Down Model}
The second part used in the ALC model is the top-down model\cite{Frandsen2006a,Calaf2010a}, which is also summarized in Shapiro et al.  \cite{Shapiro2019}. It regards the wind farm as homogenized surface roughness in the atmospheric boundary layer.  The model assumes the existence of two constant stress regions, one below the turbines and one above the turbines.  Each layer has friction velocities, $u_{*,lo}$ and $u_{*,hi}$, and roughness heights, $z_{0,lo}$ and $z_{0,hi}$, respectively.  The model connects the two regions through the wind turbine layer\cite{Calaf2010a}, with a momentum balance resulting in a relationship between the two friction velocities and the planar average velocity $\bar{u}_h$ at hub-height $z_h$ of the turbine:
\begin{equation}
    u_{*,hi}^2 = u_{*,lo}^2 + \frac{1}{2} c_{ft} \bar{u}_h^2,
\end{equation}
where $c_{ft}$ is the planform thrust coefficient that represents the momentum extracted from the flow by the turbines' axial flow resistance.  This coefficient is obtained from the wake model in implementation of the ALC model and is defined later in the paper.
\par
In the wind turbine layer, which is defined as $z_h - R \leq z \leq z_h + R$, where $R$ is the radius of the turbines, the turbulent flow is assumed to include the effects of an additional eddy viscosity $\nu_{w*}$. 
To find the eddy viscosity used in this region, the region is split in half and the total eddy viscosity depends on the friction velocities.  In the region below the hub height  $z_h - R \leq z \leq z_h$, it is represented as $(1 + \nu_{w*})\kappa u_{*,lo}$, and in the region above the hub height $z_h \leq z \leq z_h+R$, it is represented as $(1 + \nu_{w*})\kappa u_{*,hi}$.  The extra eddy viscosity term is assumed to be approximately $\nu_{w*} = 28 \sqrt{\frac{1}{2} c_{ft}}$ \cite{Calaf2010a}. 
\par
We know the lower roughness height $z_{0,lo}$ from the incoming flow, and using momentum conservation and velocity continuity,  the roughness height due to the wind farm is given by\cite{Calaf2010a}:
    \begin{equation}
    \label{eq:z0hi}
        z_{0,hi} = z_h \left(1 + \frac{R}{z_h} \right)^{\beta} \text{exp}\left( - \left[\frac{c_{ft}}{2\kappa^2} + \left(\ln \left[\frac{z_h}{z_{0,lo}} \left(  1 - \frac{R}{z_h}\right)^{\beta} \right]   \right)^{-2} \right]^{-1/2}\right),
    \end{equation}
where $\beta = \nu_{w*}/(1 + \nu_{w*})$.  Using this, we can also calculate the friction velocities in both of the layers:
\begin{equation}
\label{eq:ustarhi}
    u_{*,hi} = u_* \frac{\ln (\delta / z_{0,lo})}{\ln (\delta / z_{0,hi})},
\end{equation}
\begin{equation}
        u_{*,lo} = u_{*,hi} \frac{\ln \left( \frac{z_h}{z_{0,hi}} \left( 1 + \frac{R}{z_h} \right)^{\beta} \right)}{\ln \left( \frac{z_h}{z_{0,lo}} \left( 1 + \frac{R}{z_h} \right)^{\beta} \right)},
\end{equation}
where $u_*$ is the free-stream friction velocity and $\delta$ is the boundary layer height.  

Note that the top-down model is based on the planform-averaged momentum equation and therefore requires averaging over extended horizontal areas. In order to apply it in a more localized fashion, horizontal areas associated to each of the turbines must be defined. We follow the approach of Voronoi tesselation\cite{Shapiro2019} which naturally associates cells and their areas to each turbine $n$.  Figure \ref{fig:voronoi} shows Voronoi tessellation applied to several different configurations of turbines: a regular array, a staggered array, a circular array, and a random array.  In Voronoi tessellation, each turbine is a node, and the vertices are defined as points that are equidistant from three separate nodes, as can be most easily seen in the random array in Figure \ref{fig:voronoi}.  The edge cells are defined by using ghost points projected outside of the array, as outlined in \cite{Shapiro2019}.  We now have a way to separate the wind farm into planform areas belonging to each turbine in a way that can be generalized to any wind farm layout.   

In the ALC model the friction velocities and roughness heights can now be calculated individually in each Voronoi cell, giving information on flow conditions in the areas around each turbine. Information related to the turbulence of the local flow field is then used for each turbine and cell to determine appropriate wake expansion coefficients for the wake model portion.
\par
We use the notion of a developing internal boundary layer (IBL) over a wind farm\cite{Meneveau2012a,Boersma2018a} that depends on the streamwise position $x$. The IBL height is modeled according to:
\begin{equation}
\label{eq:deltaibl}
    \delta_{ibl}(x) = \min\left[z_h + z_{0,hi}\left( \frac{x-x_0}{z_{0,hi}} \right)^{4/5}, \,\,\delta \right],
\end{equation}
where $x-x_0$ represents the distance to the beginning of the IBL, i.e. the distance to the start of the farm directly upstream to turbine $n$ and $\delta$ is the height of the overall atmospheric boundary layer.  This implies that the friction velocities evolve as a function of $x$ through the farm, until $\delta_{ibl}$ reaches the final boundary layer height $\delta$ in the fully-developed region. Analysis of the developing wind farm internal boundary layer \cite{Meneveau2012a,stevens2016dependence} shows that the friction velocity evolves similarly to a that of boundary layer flow over a surface with a smooth-to-rough transition, i.e. the friction velocity increases sharply at the transition (the front of the wind farm) but then gradually decreases again as the IBL grows and the mean velocity gradient decreases. The analysis shows that the position-dependent friction velocity $u_*(x)$ can be obtained from Eq. \ref{eq:ustarhi} by replacing the overall $\delta$ by $\delta_{ibl}(x)$ for turbine $n$. That is to say, we will evaluate $u_{*,n}$  for each individual turbine by replacing $\delta$ in Eq. \ref{eq:ustarhi} by $\delta_{ibl}(s_{n,x})$ evaluated at the position of turbine $n$. 

When dealing with an irregular wind farm, the start of the wind farm and the IBL is challenging to define. Studies\cite{Allaerts_Meyers_2017} have shown that the aggregate effect of the wind turbines causes the internal boundary layer to start slightly in front of the first turbines in the farm. 
Figure \ref{fig:BL} shows how we model this as a collective boundary layer initialization, defined for the case of a circular wind farm. In this case the start (``trip line'') of the IBL is drawn just in front of the first set of turbines.  The streamwise distance between any given turbine and the boundary layer ``trip line'' is $x_n=x-x_0$, shown with the blue arrow for a sample turbine in the farm.  
\par
In the localized top-down model, calculations can now be performed for each turbine in its own turbine cell, enabling the localization needed to represent both the spatial non-uniformities due to arbitrary turbine placements as well as a possibly nonuniform ($y$-dependent) velocity inflow.  
For a standard boundary layer the friction velocity of a boundary layer flow over a rough surface with roughness $z_{0,lo}$ and inflow velocity $U_\infty$ (mean velocity at hub-height $z=z_h$ upstream of the wind farm) would be given by $u_* = U_\infty \kappa/\ln(z_h/z_{0,lo})$. If the inflow is $y$-dependent, and if the freestream inflow velocity corresponding to turbine $n$ is denoted as $U_{\infty,n}$ (it can be computed by averaging the inflow over the cell),  we can write $u_{*,n} = U_{\infty,n} \kappa/\ln(z_h/z_{0,lo})$. Replacing in Eq. \ref{eq:ustarhi} to obtain the actual friction velocity for turbine $n$, we can write
\begin{equation}
\label{eq:ustarhin}
    u_{*,hi,n} = U_{\infty,n}\, \frac{\kappa}{\ln(z_h/z_{0,lo})} \, \frac{\ln (\delta_{ibl}(x_n) / z_{0,lo})}{\ln (\delta_{ibl}(x_n) / z_{0,hi})}.
\end{equation}

Once the friction velocity is known, the top-down model provides a prediction for the mean velocity at hub height\cite{Calaf2010a}, and when applied individually to each cell, the mean velocity is given by  
\begin{equation}
    \bar{u}_{h,n}^{td} = \frac{u_{*,hi,n}}{\kappa} \ln \left(\frac{z_0}{z_{0,hi}} \left( 1 + \frac{R}{z_h}\right)^{\beta}  \right) 
    \label{eq:uhtd}
\end{equation}
Note that in order to evaluate $\bar{u}_{h,n}^{td}$ we require $z_{0,hi}$ which according to Eq. \ref{eq:z0hi} depends on the planform thrust coefficient $c_{ft}$. This value can differ from turbine to turbine since the total turbine forces and momentum exchanges affecting the development of the local internal boundary layer may differ across the wind farm.  In order to determine $c_{ft,n}$ for each individual turbine $n$, information from the wake model (section A. above) is required as described in section C. below presenting the coupling of both models.

\begin{figure*}[htb!]
  \begin{center}
    \includegraphics[scale=1]{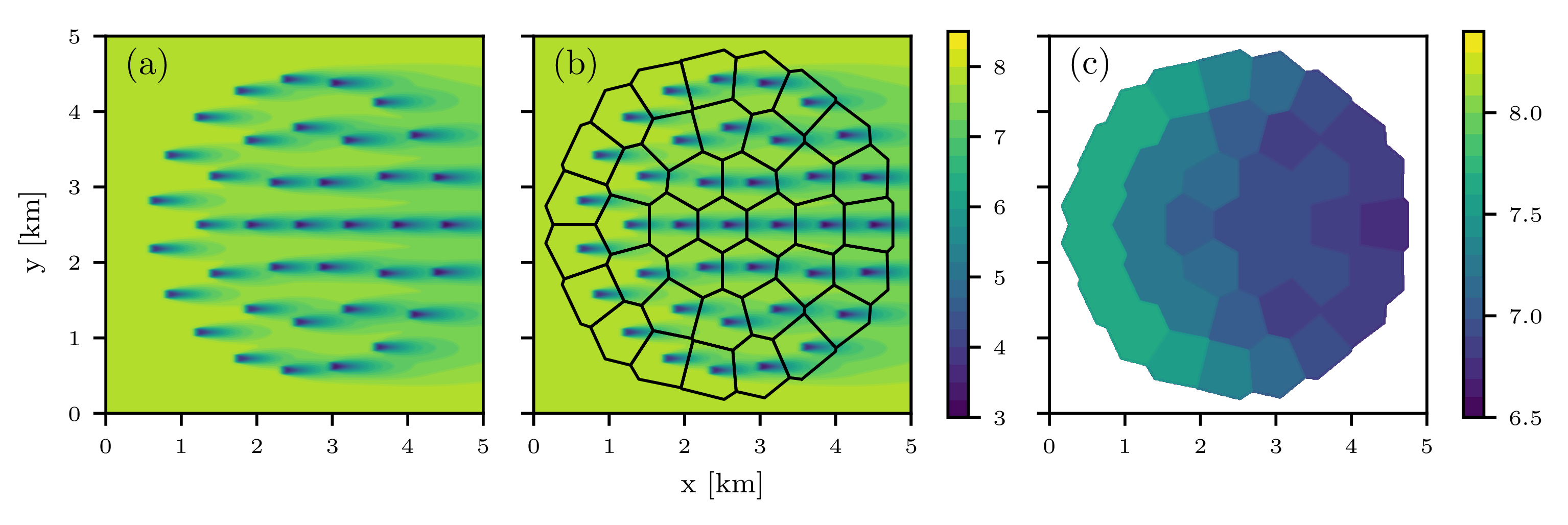}    
  \end{center}
  \caption{\label{fig:wm_Voronoi_ex} The planar velocity from the wake model is calculated by taking the (a) the velocity field given by the wake model and considering the planar average in each Voronoi cell shown in (b). The average velocity from each cell is shown in (c). (Note the use of a different color scale.) }
\end{figure*}

\subsection{Coupling of wake and top-down models: ALC Model}
The wake model and the top-down model are coupled by comparing their respective predictions of the average planar velocity values in each Voronoi cell.  In the top-down model, these are the velocities $\bar{u}_{h,n}^{td}$  calculated in each Voronoi cell according to Eq. \ref{eq:uhtd}, but the output of the wake model is a velocity field $u(x,y,z)$ from Eq. \ref{eq:velwake}.  In order to compare the wake model to the top-down model predictions, a cell-averaged wake model velocity is defined simply by averaging the velocity field predicted by Eq. \ref{eq:velwake} at $z=z_h$ over each cell for each turbine $n$. We denote this cell averaged wake model velocity as $\bar{u}_{h,n}^{wm}$. The steps are illustrated in  Figure \ref{fig:wm_Voronoi_ex}.  The average planar velocities in each cell are shown in Figure \ref{fig:wm_Voronoi_ex}(c).  Ideally one would want the velocities predicted by the top-down and wake methods to yield the same cell averaged velocities, for each turbine. However, recall that the parameter $\alpha$ required to specify the wake expansion parameter $k_w$ was left unspecified. We now follow the basic idea of the CWBL approach \cite{Stevens2015a} to determine the wake expansion using the condition that both approaches match. Since in the ALC model one has many cells, we find the $\alpha$ that minimizes the square difference between the average planar velocities from each model over all cells (the whole wind farm), according to
\begin{equation}
    \min_{\alpha} \sum_n \left(\bar{u}_{h,n}^{wm} - \bar{u}_{h,n}^{td}  \right)^2.
\end{equation}
With $\alpha$ determined, one can now calculate wake expansion coefficients to fully define the wake model using the following equation 
\begin{equation}
   k_{w,n} = \alpha \, \frac{u_{*,hi,n}}{u_{\infty,n}},
   \label{eq: k for ALC}
\end{equation}
where we define $k_{w,n}$ as the wake expansion coefficient for the $n^{th}$ turbine, using the cell-specific friction velocity found in Equation \ref{eq:ustarhin}.

\begin{figure}[h!]
  \begin{center}
    \includegraphics[scale=1]{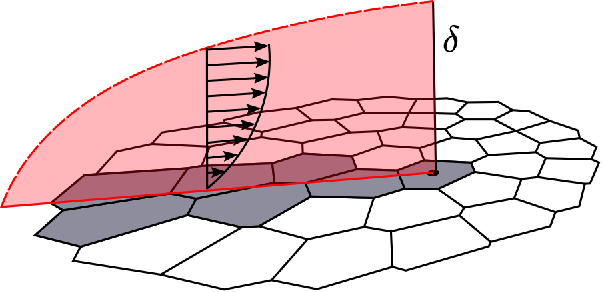}
  \end{center}
  \caption{\label{fig:delta} Illustration of how the cells are selected for inclusion in the calculation of the planform thrust coefficient, where the grey cells are included in the calculation for this particular turbine.}
\end{figure}

However, evaluation of the top-down model still requires specification of the planform thrust coefficient $c_{ft}$ in Eq. \ref{eq:z0hi}, which in turn determines the evolution of $\delta_{ibl}(x)$ as well as the top-down mean velocity from Eq. \ref{eq:uhtd}. The planform thrust coefficient $c_{ft}$ is defined as the total force per unit horizontal area. We argue that the relevant area affecting the turbulence at turbine $n$ is the area of all the Voronoi cells upstream of the turbine, as illustrated in the sketch in Figure \ref {fig:delta}. This is the region over which the IBL evolves until it reaches turbine $n$ and it is reasonable to consider the vertical momentum flux averaged over this entire ``upstream'' area to characterize the evolution of the boundary layer reaching turbine $n$.

When calculating the planform thrust coefficient, the cells between the current cell and the front of the farm are included in the calculation of total force and total momentum flux which leads to the definition of a local planform area thrust coefficient:  
\begin{equation}
    c_{ft,n} = \frac{\pi R^2 \sum\limits_{i\in line} C_{T,i}' u_{d,i}^2 }{\sum\limits_{i\in line} A_{pf,i}(\bar{u}_{h,i}^{wm})^2 }.
\end{equation}
Here $A_{pf,n}$ is the planar area of the voronoi cell for the $n^{th}$ turbine, and $line$ refers to the cells that would be crossed if a line were drawn from the current cell to the front of the farm.  This is illustrated in Figure \ref{fig:delta}, where the line is denoted in red and the cells included in the calculation are shaded.  
Moreover, $u_{d,i}$ is the wake-model velocity field $u(x,y,z)$ (obtained from Eq. \ref{eq:velwake}) averaged over the rotor disk area $A_d$ for turbine $i$:
\begin{equation}
    \label{eq:diskavu}
    u_{d,i} = \frac{1}{\pi R^2} \, \iint_{A_d} u(x=s_{i,x},y,z) \, dy dz .
\end{equation}
The coefficient $c_{ft,n}$ can be thought of as the  ratio of the total force applied by the turbines in the cells upstream and including turbine $n$ that affect the local boundary layer development, and the total horizontal momentum flux associated to the horizontally averaged mean velocity over the same planform area. The addition of the cells in front of the current cell is meant to represent the effect of the developing boundary layer over the farm and links the downstream cells to those upstream, since the turbines in the upstream cells influence those in their wakes. The coefficient $c_{ft,n}$ is then used to determine a local roughness height $z_{0,hi,n}$ that is then used in the evaluation of $\delta_{ibl}(x_n)$ and $u_{*,hi,n}$ according to Eqs. \ref{eq:deltaibl} and \ref{eq:ustarhin}. 

\begin{figure*}[htb!]
  \begin{center}
    \includegraphics[scale=0.6]{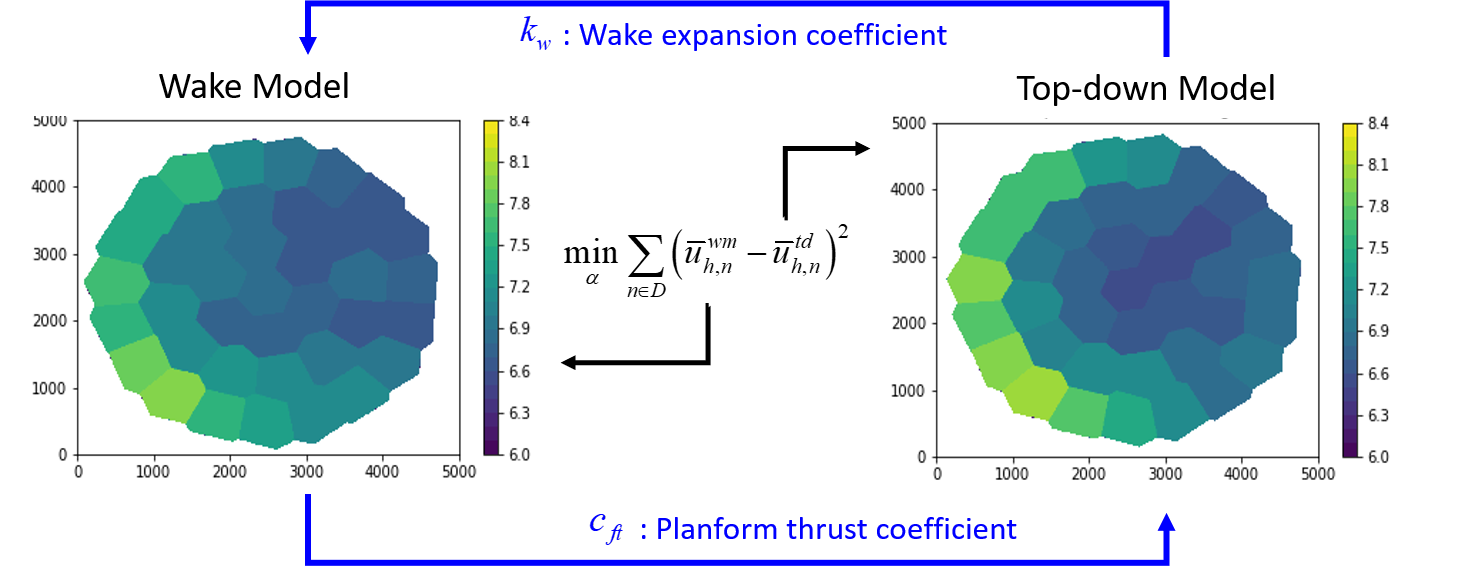}
  \end{center}
  \caption{ \label{fig:overview} A summary of the ALC model that shows how the planar velocities between the two models are compared and the difference is minimized.}
  \label{overview}
\end{figure*}

Figure \ref{fig:overview} summarizes the ALC model using an example of the planar average velocity view from each model.  Using the developing boundary layer approach for the whole farm enables the model to find one $\alpha$ value that includes input from all the cells.  The top-down model provides the wake expansion coefficient to the wake model, while the wake model provides the planform thrust coefficient to the top-down model. 

The ALC model differs from other models in that it provides significantly more information about the flow than conventional wake models that typically only furnish information about the mean flow or include entirely empirical correlations for turbulence intensities.  The ALC model also provides friction velocities for each of the cells which gives information on the turbulence conditions of the flow and can be used to quantify unsteady loading characteristics for each individual turbine.  Additionally each cell also is associated to an individual wake expansion coefficient, even though only one $\alpha$ is selected for the whole farm.  Having individual $k_w$ values for wakes emanating from each turbine can represent, for example, situations in which the expansion rate for turbines close to the entrance is lower than the wake expansion rate further downstream, where turbulence is enhanced.  

\section{\label{sec3} Model Validation For Circular Wind Farm}
\begin{figure}[h!]
  \begin{center}
    \includegraphics[scale=0.7]{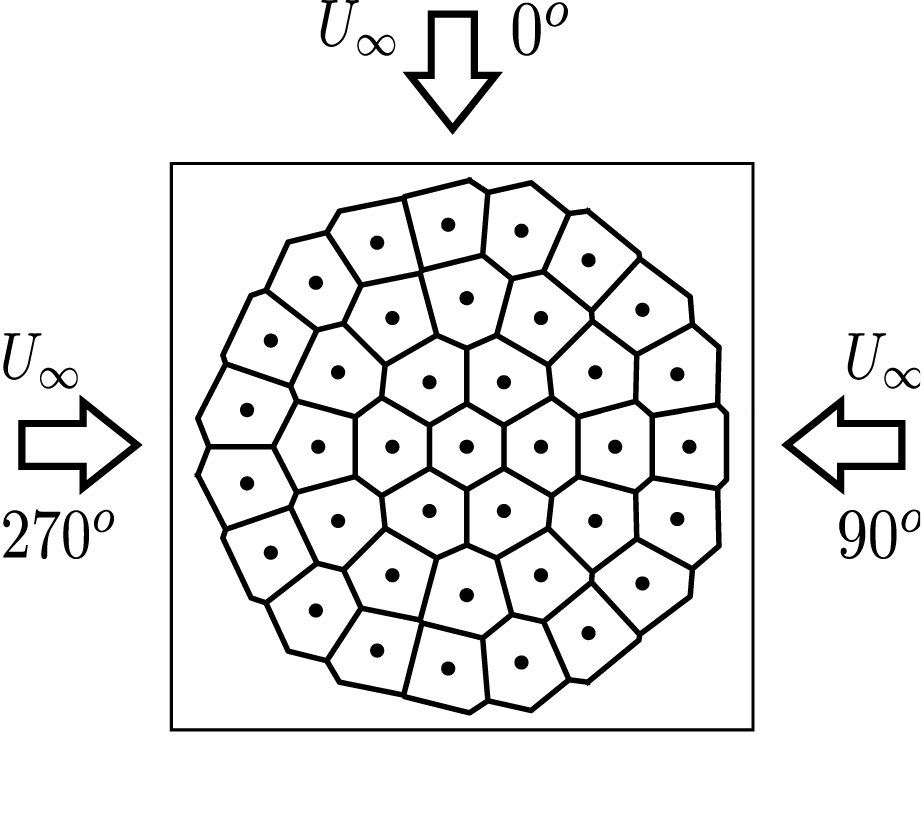}
  \end{center}
  \caption{\label{fig:orientation} Orientation of wind direction for the circular farm}
\end{figure}

The model is validated using data from LES generated at the National Renewable Energy Laboratory (NREL) for a circular wind farm configuration as shown in Figure \ref{fig:orientation}.  The circular wind farm is challenging to model due to the complex wake interactions without a regular array of turbines. The LES uses the SOWFA code \cite{SOWFA} and turbines are the NREL 5-MW reference turbine\cite{Jonkman09}. An actuator line model was used to represent the turbines in this simulation.  The wind farm in this instance is symmetric around the east-west axis.  The results from the LES are compared to the results from the ALC model in terms of total wind farm power as well as power from individual wind turbines. Comparisons are done for 12 inflow wind directions spanning 360 degrees every 30 degrees.  Figure \ref{fig:orientation} shows the orientation of the wind direction: when the wind is said to be coming from $0^o$, it is coming from the north, and then proceeds clockwise (for $90^o$ the flow is going from right to left, and for $270^o$ the flow is going left to right). 
\begin{figure}[h!]
  \begin{center}
    \includegraphics[scale=0.5]{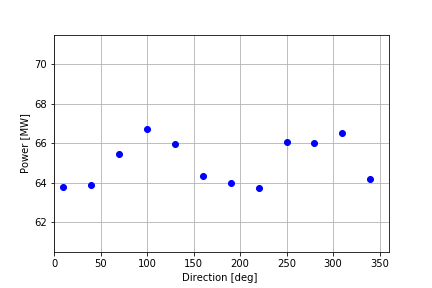}  \end{center}
  \caption{ \label{fig:sowfa_power} Average power for the circular wind farm as a function of wind direction obtained from the NREL LES.}
\end{figure}
Figure \ref{fig:sowfa_power} shows the average power as a function of wind direction obtained from LES.  The data starts at a wind direction of $10^o$. These data points were found by averaging the power measurements from the LES farm over a time interval spanning approximately 3/4 of an hour in real time of farm operation.

\subsection{Model Setup}
 To provide an accurate comparison, the  parameters used to evaluate the ALC  model are selected to closely match those of the LES.  The simulated NREL 5-MW reference turbines\cite{Jonkman09} have a hub height $z_h = 90$ meters and a diameter $D=126$ meters, and so the ALC model uses those same parameters. Additionally, we need the inflow velocity profile, the lower roughness height, and the maximum boundary layer height to evaluate the model.  The inflow velocity distribution $U_\infty(y)$ is found using average inflow velocity data from the LES.  All of the LES have very similar inflow profiles, so an average over the 12 simulations was used as the inflow profile for the model for all wind directions.  For the lower roughness height, we used the same value as was used in the LES: $z_{0,lo}=0.15$ meters.  The maximum boundary layer height was set to $\delta = 750$ meters, since the simulations had a temperature inversion layer at this height, which caps the growth of the boundary layer.  Due to the inversion layer, we can expect the boundary layer to begin before the start of the wind farm.  In this case, we defined the start of the internal boundary layer ($x_0$) to occur one turbine diameter in front of the farm.
\begin{figure}[h!]
  \begin{center}
    \includegraphics[scale=1]{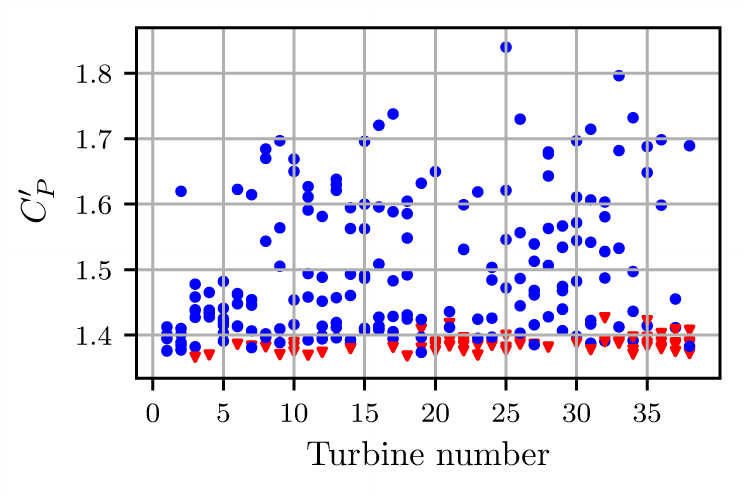}
  \end{center}
  \caption{\label{fig:Cpp} The unwaked (\color{red} $\blacktriangledown$\color{black}) and waked (\color{blue} \textbullet \color{black}) local power coefficients calculated from the data for wind directions $10^o - 190^o$ (in $30^o$ increments)}
\end{figure}
 In addition to the turbine parameters and flow conditions, we must match the thrust coefficient $C_T$ of the turbines used in the simulation and the ALC model evaluation.  This coefficient is used in the relation
 \begin{equation}
     T = \frac{1}{2} \pi R^2 \rho C_T U_{\infty}^2,
 \end{equation}
 where $R$ is the turbine radius, $\rho$ is the density of the air, and $T$ is the force between the air and the turbine.  The coefficient of power determines the power the wind turbine produces:
 \begin{align}
     P = \frac{1}{2} \pi R^2 \rho C_P U_{\infty}^3
 \end{align}
 The free-stream velocity can be related to the velocity at the disk $u_d$ using 
 \begin{equation}
     C_P \, U_{\infty}^3 = C_P^\prime \, u_d^3,
 \end{equation}
 where $C_P^\prime$ is the ``local'' power coefficient of the turbine \cite{Meyers2010a}. Note that for an ideal actuator disk, the Betz limit corresponds to $C_P^\prime=2$.  In our case, we have the power produced by each turbine from the LES as well as the velocity on the disk for the directions $10^o-190^o$ (in $30^o$ increments: $7$ directions) as data.  We can then find the local coefficient of power by rearranging the previous equation:
 \begin{equation}
     C_P^\prime = \frac{2P}{\pi R^2 u_d^3 \rho}.
 \end{equation}
  Using this expression, we can average over the time series to find an average local coefficient of power for each turbine.  Figure \ref{fig:Cpp} shows the local coefficients of power calculated for all seven directions, plotted by turbine number.  The red triangles represent unwaked (or freestream) turbines in that orientation and the blue circles represent the waked turbines.  There is significant spread in the local coefficients of power for the turbines. However, all of the freestream turbines are located at the bottom of the plot, around $C_P^\prime \approx 1.4$.  Since all of the spread occurred in the waked turbines, where turbulent fluctuations may skew calculations of the local power coefficient, the waked turbines were judged to be a less accurate measure for this quantity.  Therefore, we used the average value from the unwaked turbines, $C_P^\prime = 1.387$, for the local power coefficient in the ALC model.

Once the local power coefficient was calculated, Blade Element Momentum theory could be used to find the relationship between $C_P^\prime$ and $C_T^\prime$: 
\begin{equation}
    C_P^\prime = 0.9032 C_T^\prime
\end{equation}
where the prefactor $\gamma = 0.9032$ was determined using the specifications of the NREL 5MW turbine (we are grateful to Dr. Carl Shapiro (personal communication) for performing this calculation).  In the ALC model, $C_T^\prime$ is used to calculate the planform thrust coefficient and the local forcing of velocity deficit in the wake model, while $C_P^\prime$ is used to calculate the power from the velocity computed at the disk.
In order to compute the power from turbine $n$, we thus use
\begin{equation}
    P_{n} = \frac{1}{2} \rho C_P^\prime \pi R^2 u_{d,n}^3,
\end{equation}
where $u_{d,n}$ is the disk averaged velocity computed from the ALC model according to Eq. \ref{eq:diskavu}. 

\begin{figure*}[htb!]
  \begin{center}
    \includegraphics[scale=1]{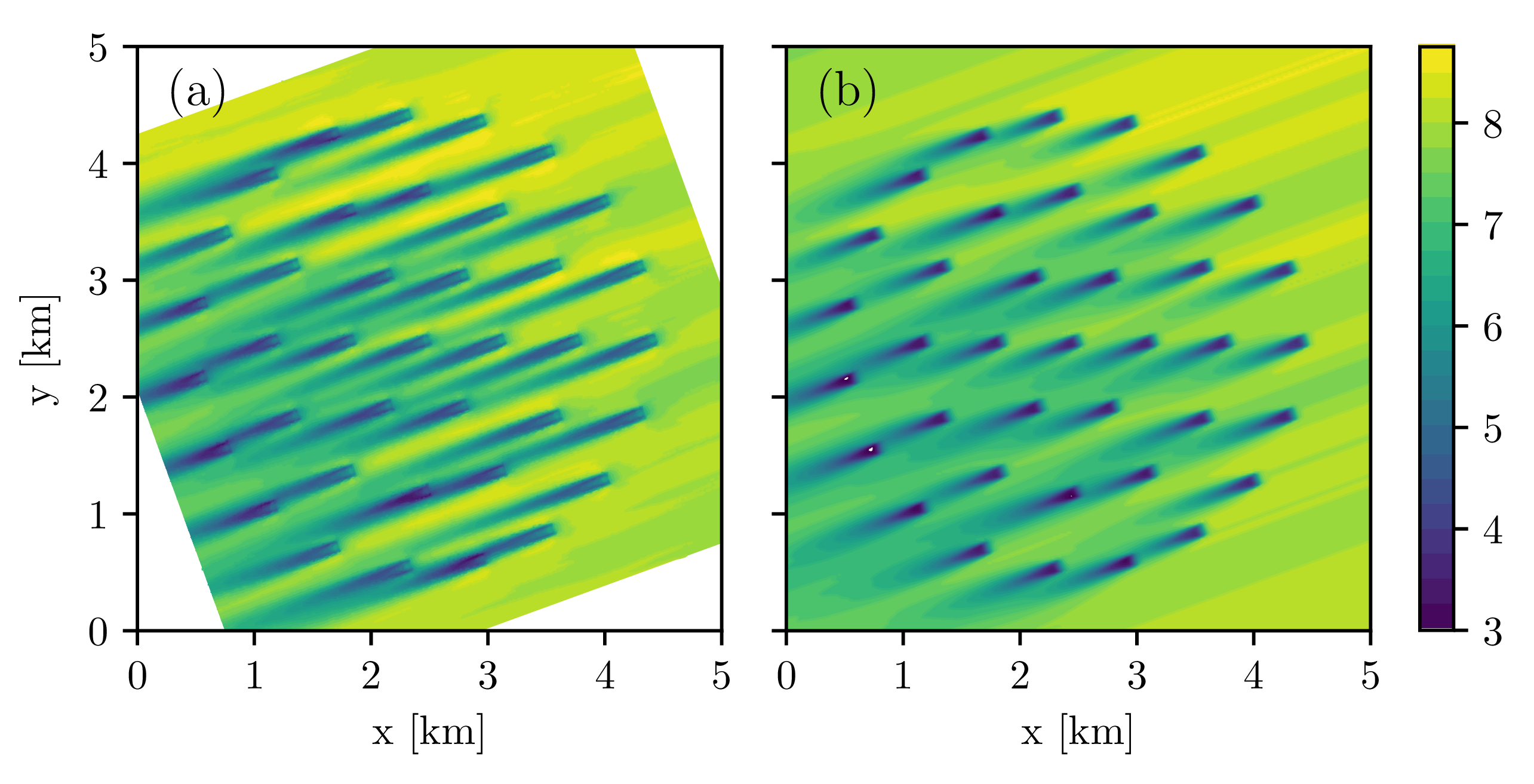}
  \end{center}
  \caption{\label{fig:70 deg vel field} Comparison of (a) the time-averaged velocity field from the LES data to (b) the velocity field calculated from the wake model for a wind direction of $70^o$}
\end{figure*}

\subsection{Results}
The ALC model was run using the parameters outlined above for the cases where the wind is approaching from $0^o - 360^o$ in increments of $5^o$.  In the directions where we have LES data, we compared with the output of the ALC model.  Figure \ref{fig:70 deg vel field} shows the average velocity field from the LES (a) and the velocity field calculated by the wake model in the ALC model (b) when the incoming wind direction is $70^o$.  We can see that with the addition of the nonuniform inflow, the wake model is able to reproduce the variations of higher and lower velocities seen across the LES wind farm, which continue back through the wind farm.  The wake model is also able to capture the wake interactions between the turbines, as can be seen by examining the wakes at the back of the farm.

\begin{figure*}[htb!]
  \begin{center}
    \includegraphics[scale=1]{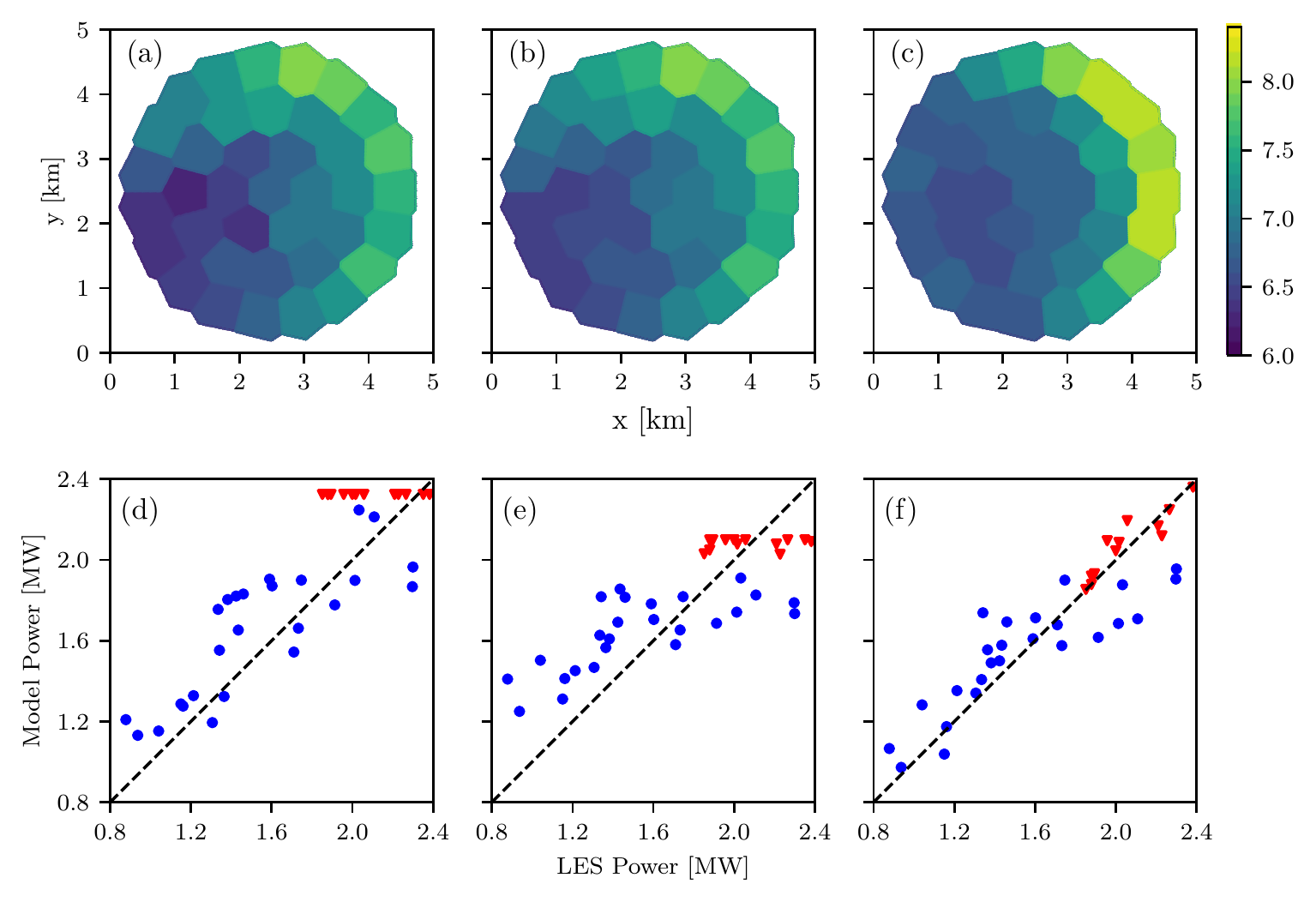}
  \end{center}
  \caption{\label{fig:70 deg} For a wind direction of $70^o$, in (a)-(c) the average planar velocities is each cell are compared from the following sources: (a) the LES data, (b) the wake model, and (c) the top-down model. The scatter plots show the power for the unwaked (\color{red} $\blacktriangledown$\color{black}) and waked (\color{blue} \textbullet \color{black}) turbines compared to the LES power from (d) the Jensen model, (e) the CWBL model from Shapiro et al\cite{Shapiro2019}, and (f) the ALC model, where a 1:1 relationship is represented by a $45^o$ line (- -)  }
\end{figure*}
For the same case of $70^o$, Figure \ref{fig:70 deg} compares the planar averaged velocities in each Voronoi cell over the farm, calculated using the method shown earlier in Figure \ref{fig:wm_Voronoi_ex} for the LES and wake model velocity fields. The top row of Figure \ref{fig:70 deg} shows the planar averaged velocities of (a) the LES, (b) the wake model and (c) the top-down model.  We can see that the  wake model and the LES average velocity fields compare well with each other.  The freestream turbines agree well, which is a result of the use of a nonuniform inflow velocity profile. Further back in the wind farm, the LES average velocity plot has a pattern of cells where some of the cells have a slower average velocity and thus more wake effects. The wake model is able to capture the overall patterns of  cell velocities well, with slightly less magnitude for some turbines (cells). 
We can also see that the wake model and the top-down model do not match exactly, as can be expected since only the mean square error is minimized in determining $\alpha$.

Next, we compare the ALC approach to prior wind farm analytical models. In the second row of Figure \ref{fig:70 deg} we compare the Jensen model, the CWBL model and the ALC model using scatter plots of the power produced by each turbine, with the LES power on the y-axis and the model power on the x-axis.  The red triangles are the freestream or unwaked turbines while the blue circles are the waked turbines.  The $45^o$ line that the models are aiming for is marked with a black dashed line.  The left plot (d) shows the scatter plot using the Jensen model with a wake expansion coefficient derived from the average friction velocity of the inflow, which was $k=0.0653$.  The middle plot (e) uses the Coupled Wake Boundary Layer model with Voronoi cells and a uniform inflow velocity \cite{Shapiro2019}.  In both of these cases, we can see that the uniform inflow prevents the unwaked turbines from capturing the behavior of the LES, causing all of the unwaked turbines to give almost the same value, falling above and below the $45^o$ line.  In the right plot (f), we show results from the ALC model.  We can see that the freestream turbines are much closer to the $45^o$  line as a result of the nonuniform inflow profile used.  
When we apply the ALC model, the waked turbines are also closer to the LES results, especially those with lower power, whose prediction is aided by the inclusion of the developing internal boundary layer and the corresponding averaged input in the planform thrust coefficient used in the top-down model.

\begin{figure}
    \begin{center}
    \includegraphics[scale=1]{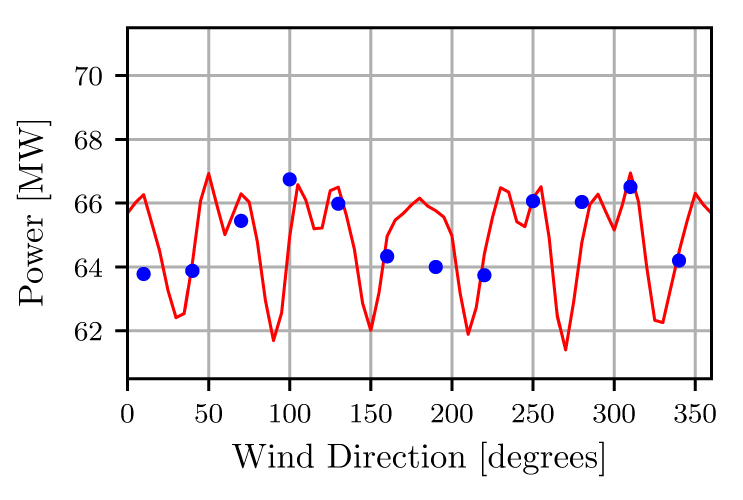}
  \end{center}
  \caption{\label{fig:power comp} Comparison of ALC model-predicted power with the model run in $5^o$ increments ({\color{red} --}) to the power obtained from LES ({\color{blue} \textbullet} for the circular wind farm. The LES results are repeated from Figure \ref{fig:sowfa_power}. }
\end{figure}

\begin{figure}[h!]
  \begin{center}
    \includegraphics[scale=1]{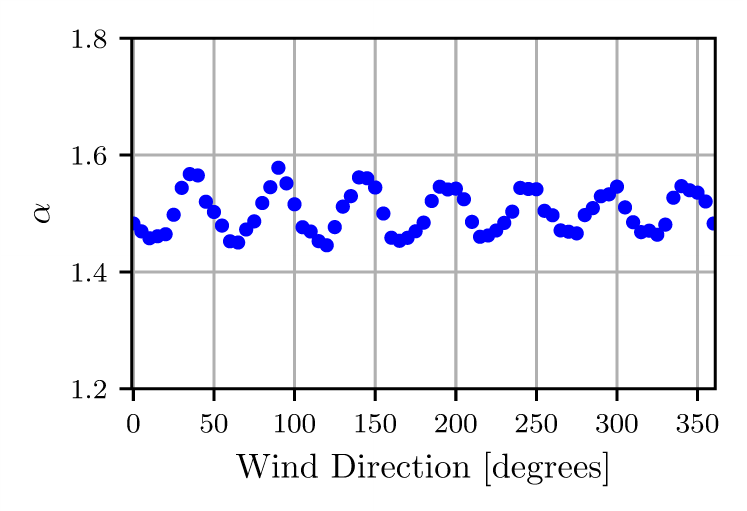}
  \end{center}
  \caption{\label{fig:alpha values} Model parameter $\alpha$ obtained from the error minimization in the ALC model as a function of wind direction for the circular wind farm.}
\end{figure}

In Figure \ref{fig:alpha values} we show the $\alpha$ parameter obtained from the error minimization over the entire farm, as function of wind direction. The fact that the results are on the order of unity (between 1.45 and 1.6) confirms the validity of the scaling assumptions underlying Eq. \ref{eq: k for ALC}. A non-trivial dependence on angle can be observed, a result of the complicated relationships involved in the model, the turbine geometric layout and the transverse variations in inflow velocity.  

Next, the ALC model was run for the circular wind farm for all wind directions in increments of $5^o$.  Figure \ref{fig:power comp} shows the results for the model compared to the LES results in 12 directions.  The model results are shown by the red line and the blue circles represent the power from the LES.  The farm is top-bottom symmetric, but the model results, while showing very similar trends, are not quite symmetric, as a result of the nonuniform inflow used for the farm.  We can see that in most cases the ALC model results match those of the LES, i.e. they fall reasonably close to the blue symbols. The two noticeable outliers are for the directions $10^o$ and $190^o$.  In both cases the model over-predicts the power, however, the results for a few degrees to the right provide a better match.  This highlights how small changes in angle can significantly impact the power output of the wind farm. Such changes are especially apparent, for example, over the range $90^o$-$110^o$.
\begin{figure*}[htb!]
  \begin{center}
    \includegraphics[scale = 1]{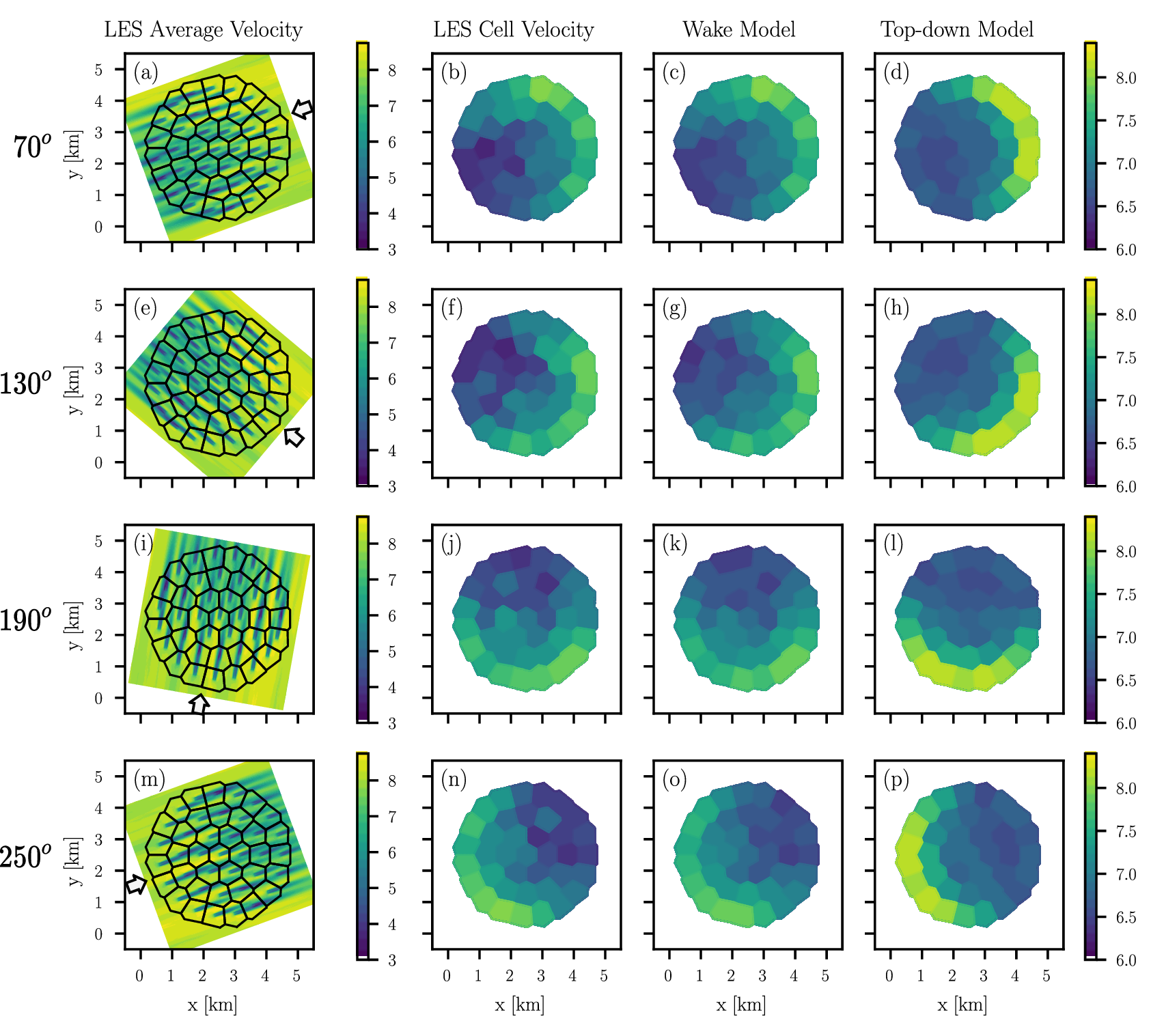}
  \end{center}
  \caption{\label{fig: planar vel comp} Comparison of LES data and the planar velocities given by the ALC model.  For $70^o$, (a) shows the average velocity data from the LES with the Voronoi cells drawn, (b) shows the average planar velocity calculated from (a), (c) shows the wake model average planar velocity, and (d) shows the same quantity from the top-down model.  The subfigures (e)-(h) show these same quantities for $130^o$, subfigures (i)-(l) show the velocites for $190^o$, and subfigures (m)-(p) show the velocites for $250^o$.}
\end{figure*}

\begin{figure*}[htb!]
  \begin{center}
    \includegraphics[scale=1]{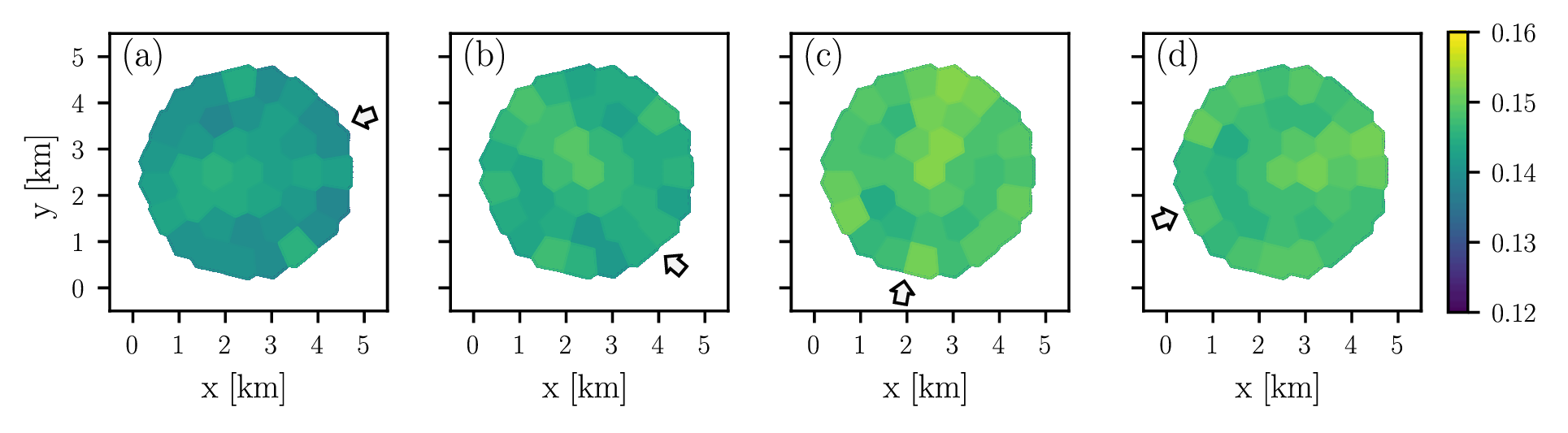}
  \end{center}
  \caption{\label{fig:k values}Contour plots shows the value of the wake expansion coefficient calculated by the model for (a) $70^o$, (b) $130^o$, (c) $190^o$ and (d) $250^o$.}
\end{figure*}

\begin{figure*}[h!]
  \begin{center}
    \includegraphics[scale=1.2]{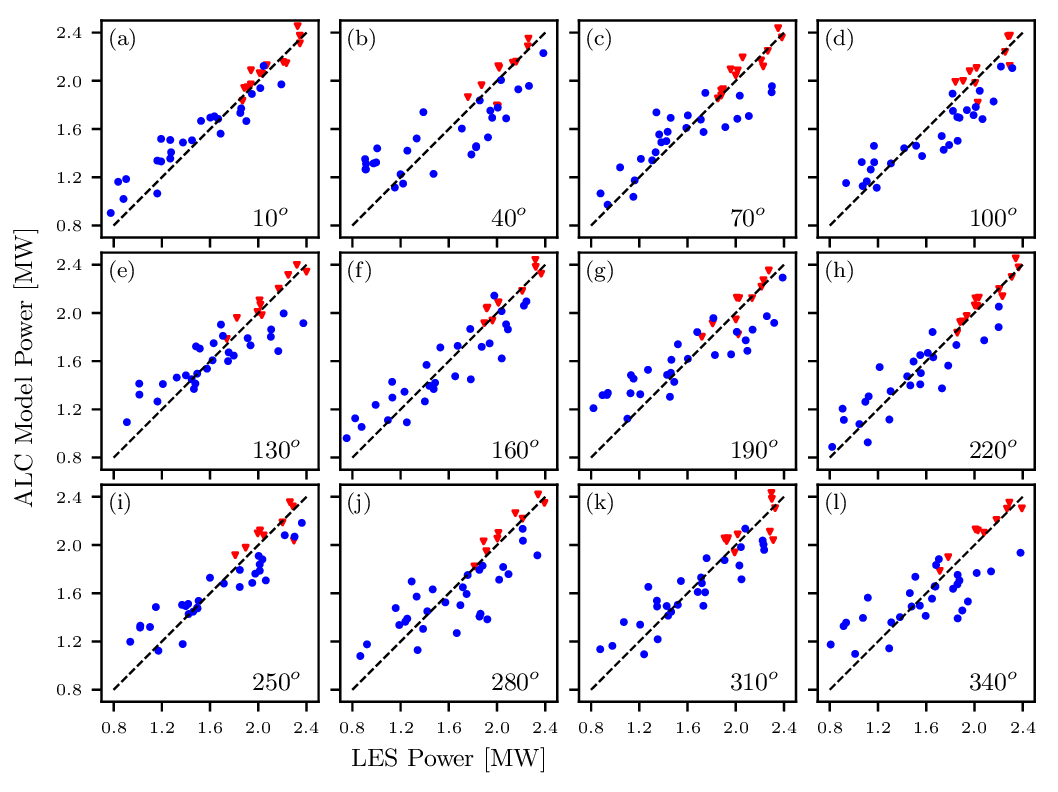}
  \end{center}
  \caption{\label{fig:full scatter}Scatter plots comparing the power predicted by the ALC model with the LES data for all the LES wind directions, where (\color{red} $\blacktriangledown$\color{black}) represents the unwaked turbines and (\color{blue} \textbullet \color{black}) represents the waked turbines and a 1:1 relationship is represented by a $45^o$ line (- -).}
\end{figure*}
Figure \ref{fig: planar vel comp} provides a more detailed comparison of the LES field to the model over four different incoming wind directions: $70^o$, $130^o$, $190^o$, and $250^o$.  Each direction corresponds to a row in the figure, while each column depicts a different quantity for that direction.  The first column shows the time averaged LES velocity field with the Voronoi cells drawn over it to show the area over which the planar velocity values are calculated.  The first column also has arrows to indicate the direction of the incoming wind.  The second column shows the cell average LES planar velocities calculated from the time average LES velocity fields.  The third column shows the average planar velocities from the wake model of the ALC model, and the last column shows the same quantity from the top-down model.  The $130^o$ and $250^o$ degree directions match quite well with the model, as can be seen from Figure \ref{fig:power comp}.  However, in the $190^o$ direction, the model overpredicts the power.  In all the cases, the wake model is able to predict the pattern of darker and lighter cells in the waked region, the back half of the wind farm, of the LES farm quite well.  While the magnitude of the velocities from the wake model are slightly higher than that of the LES, the pattern it predicts matches well. This is important in seeing how the turbines and wakes interact with each other.  The relationships between the Voronoi cells can give insight into the more complicated wake interactions.

We can further examine wake interactions between turbines by examining the wake expansion coefficient determined as part of the ALC model.  Even though only one parameter is matched across the whole farm ($\alpha$), the model gives varying wake expansion coefficients across the farm because each cell has a different freestream velocity and friction velocity $u_{*,hi,n}$.  Figure \ref{fig:k values} shows how the wake expansion coefficient differs across the farm for the same four wind directions shown in Figure \ref{fig: planar vel comp}, this time with the wind direction changing across the row.  A lighter value indicates a higher wake expansion coefficient, and thus a wider wake, more expansion and velocity recovery, and more turbulence in the flow.  A darker value indicates a lower value and a stronger wake effect and velocity deficit on the following turbines. Here, we can see that the wake expansion coefficient is lower overall for the first direction (a) $70^o$, but is higher for the later two directions: (c) $190^o$ and (d) $250^o$.  A common feature in all directions, however, is that the wake expansion coefficient is the typically highest in the back center of the farm, as referenced from the freestream wind direction.  This trend reproduces the realistic effect that an increase of turbulence further back in the farm is due to cumulative wake interactions.

In addition to the planar velocity comparisons presented above, we can also examine a more quantitative view of the model performance using scatter plots for all the directions.  Figure \ref{fig:full scatter} shows the scatter plots for all 12 directions provided by the LES data.  The unwaked turbines are represented by the red triangles and the waked turbines are represented by the blue circles.  The $45^o$ line with LES data is represented by the black dashed lines.  We can see that in each case, the ALC model enables the power predictions for the freestream turbines to reflect the actual behavior of the freestream turbines in the the LES results.  The results for the waked turbines have more spread overall, but follow the trend of the line quite well.  The extra spread of the waked turbines could be explained by noting the spread in the local coefficients of power calculated from the waked turbines, which is evident in Figure \ref{fig:Cpp}, shown earlier.  Since we chose the coefficient of power to use for the model based on the unwaked turbines, the waked turbines could deviate from this constant value in the LES somewhat, possibly due to wake interactions, causing more spread in the results.

\section{\label{sec4} Model Validation Using a Mixed Regular and Nonuniform Wind Farm}

The model was also compared with LES data from a simulation using a wind farm layout that mixes a regular array region with a non-uniform region of the wind farm.  In this case, the wind farm starts with and array consisting of four staggered rows, and then has a total of fourteen additional turbines placed in a random fashion behind the staggered-array turbines.  In this simulation, an actuator disk model was used to represent the turbines, using the JHU LESGO code \cite{lesgo} with an actuator disk correction \cite{Shapiro2019a}. The code has been amply validated on wind energy applications  
\cite{lignarolo2016validation,Stevens2018a,martinez2018comparison}. The simulations use as inflow condition a field generated by the concurrent-precursor approach \cite{Stevens2014a}. The turbines have a diameter and hub height of $D=z_h=100$ meters and a ground surface roughness height of $z_{0,lo}=0.1$ meters is used to prescribe the bottom boundary condition of the LES. The main simulation domain uses $(N_x \times N_y \times N_Z = 384 \times 256 \times 128$) grid-points and the Lagrangian scale dependent dynamic subgrid-scale model is used.  The coefficient of thrust for these turbines was kept constant at $C_T^\prime=1.33$ throughout the simulation. All these parameters were also used in evaluating the ALC model. The LES was run for a single inflow direction of $270^o$, giving only one direction of reference for comparison to the model.  The LES was run for a time period of approximately ten flow-through times for the farm, which translates to roughly $1.75$ hours in real time, also providing an adequately averaged inflow condition for the model.

\begin{figure*}[h!]
  \begin{center}
    \includegraphics[scale=1]{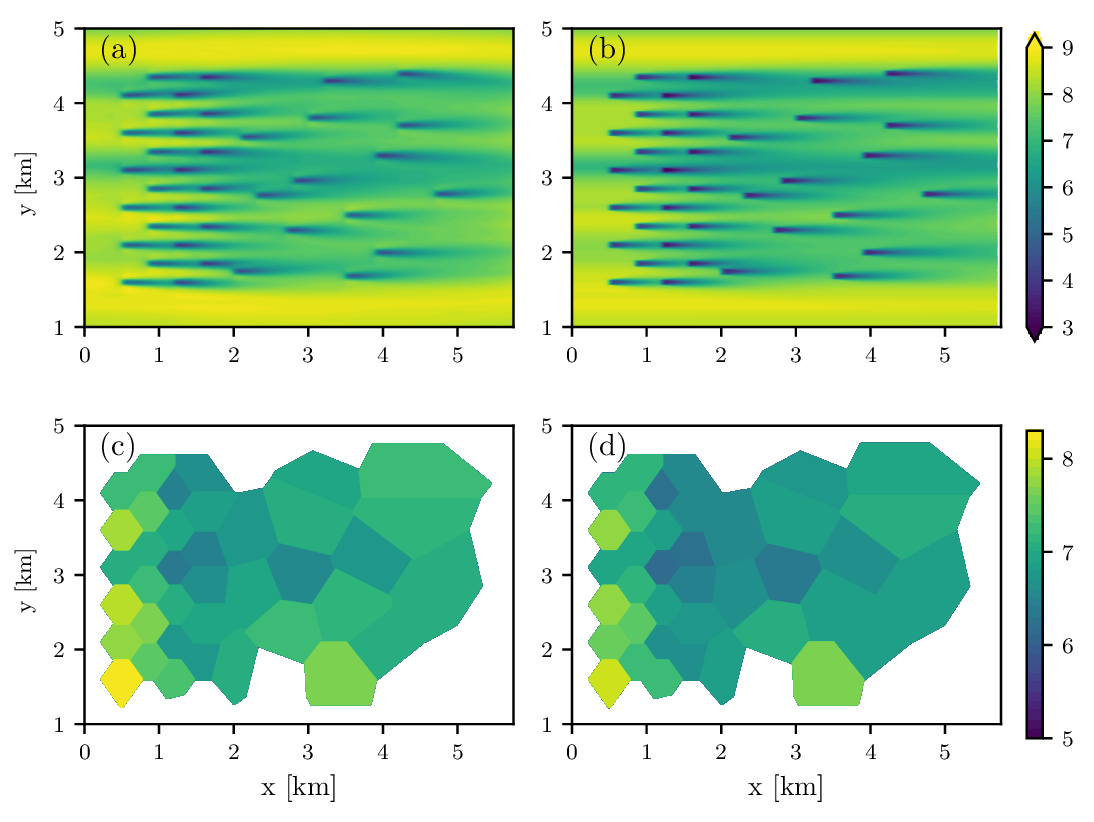}
  \end{center}
  \caption{\label{fig:random_vel} Comparison of the velocity fields (a) of the LES  with (b) the ALC wake model, and the comparison of the hub height average cell velocity from (c) the LES and (d) the ALC model.}
\end{figure*}

Figure \ref{fig:random_vel} shows the comparison of the time averaged streamwsie velocity field at hub-height given by (a) the LES and (b) the ALC model.  The inflow used in the ALC model was taken from the time-average LES velocity field at $x=0$ in front of the turbines.  The dimensionless wake expansion parameter found through the error minimization was $\alpha = 1.76$, which is slightly higher than any of the values found in the circular wind farm study, but still of order unity.  We can see that the ALC model captures the structures of the flow well, and represents the wakes, particularly in the rear of the farm, quite accurately.  The wakes in the ALC model appear sharper and the velocity is slightly lower in the wake as compared to the LES results, but their relative speed of decay is similar in both cases.  The comparison also shows the importance of the nonuniform inflow condition to accurately model  the overall flow especially near the inlet.  The lower panels of Figure \ref{fig:random_vel} show the comparison between the average velocities in each of the VOronoi cells where (c) shows the time and cell-averaged field from the LES the (d) the mean velocity obtained from the ALC model (the wake model part).  We can see that the cell velocities match well, showing the same general trends in both cases.  The LES results display slightly higher velocities throughout the farm, which is reflected in the darker wakes from the wake model, as well.  However, the overall patterns of slower and faster velocity cells are captured well by the ALC model.

\begin{figure}[htb!]
  \begin{center}
    \includegraphics[scale=1]{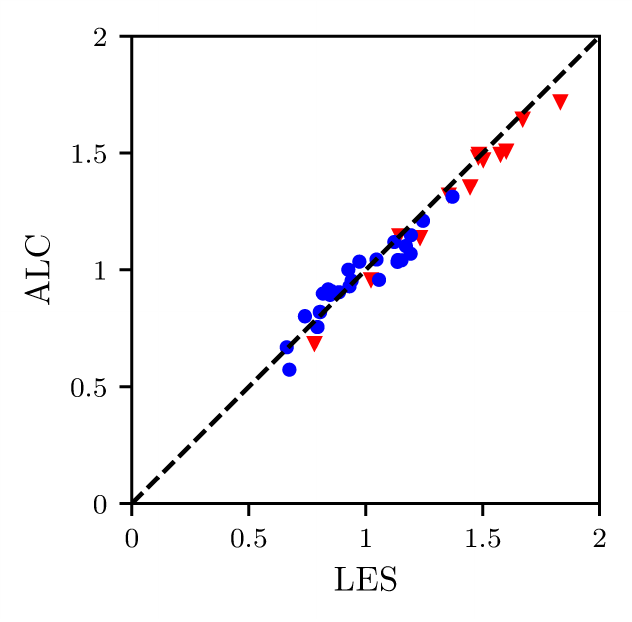}
  \end{center}
  \caption{\label{fig:scatter random} Scatter plot comparing the power predicted by the ALC model with  LES data from the mixed regular nonuniform wind farm, where (\color{red} $\blacktriangledown$\color{black}) represents the unwaked turbines and (\color{blue} \textbullet \color{black}) represents the waked turbines and a 1:1 relationship is represented by a $45^o$ line (- -).}
\end{figure}

Finally, as a more quantitative comparison between the ALC model results and the LES results, Figure \ref{fig:scatter random} shows a scatter plot comparing the power from each turbine.  As before, the unwaked turbines (first two rows at $x = 0.5$km and 1km) are denoted by the red triangles while the waked turbines further downstream are denoted by the blue circles.  We can see that the points follow the $45^o$ line rather closely, implying very good agreement of the ALC predictions with the LES results. There is substantially less scatter compared to the circular wind farm case considered in the last section. This could be due to the longer time period used for averaging which for the mixed regular-nonuniform wind farm was 1.75 hours (approximately 11 wind farm flow through times), while it was only 45 mins (approximately 4.5 wind farm flow through times) for the circular wind farm. We recall that the ALC model is designed to predict the mean flow which requires significant time averaging. In summary, we can conclude that the  ALC model was able to capture the mean flow and power production of the mixed regular-nonuniform wind farm quite well.

\section{\label{Sec5} Conclusions}
The Area Localized Coupled (ALC) model combines a wake model and a top-down model in a localized fashion to create a more generally applicable model that predicts several useful qualities about wind farms.  The model uses Voronoi cells to divide the wind farm up into areas that belong to each turbine.  Since the calculations can be applied to each cell, the model can be applied to regular as well as irregularly arranged wind farms.  The model includes an important parameter, the wake expansion coefficient, that is obtained as function of the local turbulence properties as described by the ratio of the friction velocity to the mean advection velocity. The dependence of the wake expansion coefficient on the local friction velocity determined from the growth of a local internal boundary layer enables a localized result involving different position-dependent wake expansion coefficients across the wind farm. A parameter $\alpha$ is selected to minimize the difference in the cell-averaged mean velocity predictions between the top-down and wake superposition models.  In addition, the model can be implemented using a nonuniform inflow for both  the wake and top-down model parts. This feature improves the performance of the model for all of the turbines but especially for turbines near the inflow that are directly exposed to the position-dependent inflow velocity.  

The model was validated using the LES data for two different wind farms. The first was for a circular wind farm.  After the flow conditions were matched, the model was able to accurately predict the cells of higher and lower velocity across the wind farm, calculated from the time average LES data.  The ALC model also reproduced the patterns in the cells further back in the farm, where the wake interactions are most important. The ALC model also captured general trends in the variation of total wind farm power with wind direction as obtained from LES data. It also predicted the  power of individual turbines well.  The ALC model was also applied to a mixed regular-nonuniform wind farm for which LES using an actuator disk model generated an additional dataset. The wind farm started with regular staggered rows and then continued with a set of randomly placed turbines downstream.  The velocity field provided by the wake model matched well with the time-average velocity from the LES  and the model was able to predict the power of individual turbines quite accurately.
\par
The new proposed ALC  model provides more information than typical wake models since physically relevant information about position-dependent friction velocity, internal boundary layer and roughness height becomes available. Since the Voronoi tesselation is generally applicable to any turbine arrangement, there are no limitations to apply the ALC to any wind farm geometry, regular, random or mixed.  It also handles position-dependent inflow velocities.  Additionally, the model also has the potential to be extended to a dynamic, time dependent model if instead of using the steady state solution of the partial differential equation for the velocity deficit one uses the full time-dependent solution. The top-down model still is based on a quasi-steady description of the boundary layer and so further improvements to include time-dependence in the top-down model may still be required. At any rate such possible extensions make the ALC model an attractive and versatile option to be used in wind farm design and control. Future work will include the validation of this model on operational wind farms and possible extensions of this framework to a dynamic model.

\section{Acknowledgements}
G.M.S., C.M. and D.F.G. gratefully acknowledge partial funding support from the National Science Foundation ((grant numbers DGE-1746891, CBET-1949778 and CMMI-1635430) and the Maryland Advanced Research Computing Center for computing resources.

This work was authored by the National Renewable Energy Laboratory,
operated by Alliance for Sustainable Energy, LLC, for the U.S. Department
of Energy (DOE) under Contract No. DE-AC36-08GO28308. Funding
provided by the U.S. Department of Energy Office of Energy Efficiency and
Renewable Energy Wind Energy Technologies Office. The views expressed
in the article do not necessarily represent the views of the DOE or the
U.S. Government. The U.S. Government retains and the publisher, by
accepting the article for publication, acknowledges that the U.S. Government retains a nonexclusive, paid-up, irrevocable, worldwide license to publish or reproduce the published form of this work, or allow others to do so, for U.S. Government purposes.

\end{document}